\documentstyle[aps,prl]{revtex} 
\begin{document}
\draft
\twocolumn[\hsize\textwidth\columnwidth\hsize\csname 
@twocolumnfalse\endcsname

 %Inserted by TeXtelmExtel

%\setlength{\leftmargin}{2cm}

 %Inserted by TeXtelmExtel

 %Inserted by TeXtelmExtel

 %Inserted by TeXtelmExtel

\title{On the estimate of the spin-gap in quasi-1D Heisenberg
antiferromagnets from
nuclear spin-lattice relaxation}

 %Inserted by TeXtelmExtel

\author{R. Melzi and P. Carretta}
\address{
Department of Physics "A. Volta", Unit\'a INFM di Pavia,
27100 Pavia, ITALY}

 %Inserted by TeXtelmExtel

\date{June 29, 1998}
\maketitle
%%%%%%%%%%%%%%%%%%%%
\widetext
%%%%%%%%%%%%%%%%%%%%
% \vskip -1cm
\begin{abstract}

 %Inserted by TeXtelmExtel

 %Inserted by TeXtelmExtel

 %Inserted by TeXtelmExtel

 %Inserted by TeXtelmExtel

 %Inserted by TeXtelmExtel

 %Inserted by TeXtelmExtel

 %Inserted by TeXtelmExtel

 %Inserted by TeXtelmExtel

%%%%%%%%%%%%%%%%%%%%
% \leftskip 54.8pt
% \rightskip 54.8pt
%%%%%%%%%%%%%%%%%%%%

 %Inserted by TeXtelmExtel

 %Inserted by TeXtelmExtel

 %Inserted by TeXtelmExtel

 %Inserted by TeXtelmExtel

 %Inserted by TeXtelmExtel

 %Inserted by TeXtelmExtel

 %Inserted by TeXtelmExtel

 %Inserted by TeXtelmExtel

We present a careful analysis of the temperature dependence of the nuclear
spin-lattice relaxation rate $1/T_1$ in gapped quasi-1D Heisenberg
antiferromagnets. It is found that in order to estimate the value of the gap correctly
from $1/T_1$ the peculiar features of the dispersion curve for the triplet
excitations must be taken into account. The temperature dependence of $1/T_1$ 
due to
two-magnon processes, is reported for different values of the ratio $r=J_{\perp}/J_{\parallel}$
between the superexchange constants in a 2-leg-ladder. 
As an illustrative example we compare our results to the experimental
findings for $^{63}$Cu $1/T_1$ in the dimerized chains
and 2-leg-ladders contained in Sr$_{14}$Cu$_{24}$O$_{41}$.
\end{abstract}
\newpage
%%%%%%%%%%%%%%%%%%%%
% \leftskip 54.8pt
% \rightskip 54.8pt
%%%%%%%%%%%%%%%%%%%%
]
%%%%%%%%%%%%%%%%%%%%
\narrowtext
%%%%%%%%%%%%%%%%%%%%
%\twocolumn

 %Inserted by TeXtelmExtel

 %Inserted by TeXtelmExtel

The many peculiar aspects of quasi one-dimensional quantum Heisenberg antiferromagnets (1DQHAF)
have stimulated an intense research activity during the last decade \cite{Dagotto}. Moreover, the recent
observation of superconductivity in the 2-leg-ladder compound 
(Sr,Ca)$_{14}$Cu$_{24}$O$_{41}$ \cite{Uehara} and the occurrence of a 
phase separation in high temperature superconductors (HTSC) in hole-rich and hole-depleted regions 
analogous to spin-ladders \cite{Tranquada}, have brought to a renewed interest on 1DQHAF. One of the
relevant issues is wether the spin-gap observed in some of these 1DQHAF is related to 
the one observed in the normal state of HTSC \cite{Car0}. For these reasons many NMR groups
working on HTSC have focused their attention on these systems and on the determination
of the spin-gap values in pure and hole-doped compounds
\cite{Azuma,Taki1,Taki2,Kumagai,Magishi,Car1,Car2,Mayaf,Tedo,Imai,Greg,Furuk}. However, since the early measurements,
a clear discrepancy between the values for the gap ($\Delta$) estimated by means of nuclear spin-lattice
relaxation ($1/T_1$) and susceptibility
(or Knight shift) measurements has emerged \cite{Azuma}. In many 
compounds the gap estimated by means of $1/T_1$
using the activated form $1/T_1\propto exp(-\Delta/T)$ derived by Troyer et al. \cite{Troyer}, 
turned out to be $\simeq 1.5$ times larger than the one estimated by using susceptibility or
inelastic neutron scattering measurements (see Tab. 1). Many attempt models, theoretical 
\cite{Damle} or phenomenological \cite{Imai}, have tryed
to explain these differences, however, while they were able to describe the findings
for some compounds they were not able to explain the results obtained in other gapped
1DQHAF.
In fact, as can be observed in Tab. 1, while for certain 2-leg-ladders \cite{Greg}
an agreement
between the gap estimated from $T_1$ and through other techniques is found, in several
other systems it is not \cite{Azuma,Taki2,Kumagai,Magishi,Car1,Car2,Mayaf}. It is interesting
to observe that the 1DQHAF where the agreement is observed are the ones
in the strong coupling
limit, namely either dimerized chains or 2-leg-ladders with a 
superexchange coupling along the rungs much larger 
than the one along the chains. 
Therefore, one can conclude that the disagreement is not always present
and has to be associated 
with the peculiar properties of the spin excitations in each system, i.e.
with the form of the dispersion curve for the triplet excitations. 
In this manuscript we will show that the discrepancy relies essentially on the use for $1/T_1$
of an expression which is valid in general only at very low temperatures ($T\lesssim 0.2\Delta$)
and its application to higher temperatures depends on the form of the dispersion curve for
the triplet spin excitations. In particular, for dimerized chains the  validity of 
a simple activated expression extends to higher temperatures than for a 2-leg-ladder.
As an illustrative example we will
analyse the temperature dependence of $1/T_1$ for the $^{63}$Cu nuclei in the dimer
chains (Cu(1)) and in the 2-leg-ladders (Cu(2)) contained in Sr$_{14}$Cu$_{24}$O$_{41}$
\cite{Strut}.

 %Inserted by TeXtelmExtel

 %Inserted by TeXtelmExtel

In the following we will consider the contribution to  nuclear relaxation
arising from 2-magnon Raman processes only. Namely, we will assume that 
although the system is not in the very
low temperature limit ($T\ll\Delta$), the temperature is low enough ($T\lesssim\Delta$)
so that 3-magnon
processes as well as the spin damping can be neglected.
If the large value of the gap derived by means of $1/T_1$ was due to 
these contributions, which are proportional to $exp(-2\Delta/T)$, one should observe some
discrepancy also for the 1DQHAF in the strong coupling limit, 
at variance with the experimental findings (see Tab. 1).
The approach we use follows exactly the same steps outlined in the paper
by Troyer et al. \cite{Troyer} where, by assuming a quadratic dispersion for
the triplet excitations (valid for $T\ll\Delta$), namely
\begin{equation}
E(k_x)=1+ \alpha(k-\pi)^2
\end{equation}
in units of $\Delta$, they found that
\begin{equation}
1/T_1={3\gamma^2A_o^2\over4\alpha\pi^2}{\hbar\over k_B\Delta}e^{\omega_o/2T}
e^{-\Delta/T}(0.80908 - ln(\omega_o/T))
\end{equation}
with $\omega_o$ the resonance frequency and $A_o$ the hyperfine coupling constant.
We remark that there is a factor 4 difference with respect to the equation reported
by Troyer et al. \cite{Troyer}, which is related to a different definition of the hyperfine 
hamiltonian and of
the dispersion curve. The values of the hyperfine constants are 
$A_o=120$ kOe for $^{63}$Cu(2) and $A_o=29$ 
kOe for $^{63}$Cu(1) \cite{Car1,Taki2}.
In the case of a general form for the dispersion relation, by considering that the low-energy
processes are the ones corresponding to an exchanged momentum $q\simeq 0$ and $q\simeq -2k_x$,
one can write the contribution related to 2-magnon Raman processes in the form \cite{Troyer}
\begin{equation}
1/T_1={3\gamma^2A_o^2\over\pi^2}{\hbar\over k_B\Delta}\int_0^{\pi} dk_x {e^{-E(k_x)/T}\over
\sqrt{v^2(k_x) + 2\omega_o {\partial v(k_x)\over \partial k_x}}}
\end{equation}
where $E(k_x)$ is the dispersion relation for the triplet spin excitations, normalized to the
gap value, whereas $v(k_x)=\partial E(k_x)/\partial k_x$. 
For a 2-leg-ladder a general form describing $E(k_x)$ is
\begin{equation}
E(k_x)^2=E(k_x=0)^2cos^2({k_x\over 2})+ sin^2({k_x\over 2}) + c_osin^2({k_x})  
\end{equation}
which is strongly dependent on the ratio $r=J_{\perp}/J_{\parallel}$ between the superexchange
coupling along the rungs and along the legs. 
We have taken the dispersion curves derived by Oitmaa et al. \cite{Oitmaa} from
an extensive series studies and estimated the parameters
$E(k_x=0)$ and $c_o$ accordingly. Then, starting from Eqs. 3 and 4, by means of a numerical
integration one can derive directly $1/T_1$ for a 2-leg-ladder for different values of $r$.
It should be remarked that for $r$ of the order of unity the
dispersion curve for the triplet excitations has a maximum around a wave-vector 
$k_m$ \cite{Oitmaa} (see Fig. 1) and
also low-energy processes from $k_m-k_x$ to $k_m + k_x$ could contribute to the relaxation.
However, this processes should become relevant only at $T\gtrsim \Delta$, where also 
3-magnon processes and the damping of the spin excitations become relevant.  

 %Inserted by TeXtelmExtel

 %Inserted by TeXtelmExtel

In Fig. 2 we report the results obtained on the basis of Eqs. 3 and 4 for 
$^{63}$Cu(2) for different values of the superexchange anisotropy $r$. 
One observes that while for the
dimerized chains, corresponding to the limit $r\gg 1$, $1/T_1$ follows an activated
behavior as the one given in Eq. 2, for the 2-leg-ladders 
with $r\sim 1$ one observes some differences
with respect to the simple activated behavior already at temperatures 
$T\lesssim \Delta/4$.
This analysis points out that for a 2-leg-ladder with $r$ of the order of unity it is not correct
to estimate the gap from $1/T_1$ by using Eq. 2, at least for 
$T\gtrsim \Delta/4$. In fact,
it is noticed that the quadratic approximation for the dispersion curve becomes valid for
a more restricted range of $k_x$ around $\pi/a$ as $r$ decreases (see Fig. 1). 
This seems to contradict the results reported in Fig. 2a, where the departure from the quadratic
approximation is found more pronounced for $r=1$ than for $r=0.5$. However, this artifact is
related to the choice of the horizontal scale, namely to have reported $1/T_1$ vs. $\Delta/T$,
since $\Delta$ increases with $r$ \cite{Dagotto}. In fact, if we report 
$1/T_1$ vs $J_{\parallel}/T$ (Fig. 2b), 
with $J_{\parallel}$ independent of $r$, one immediately notices
that the deviation from the quadratic approximation starts at lower temperatures for
the lowest value of $r$.

 %Inserted by TeXtelmExtel

One can then analyse the experimental data on the basis of Eq. 3 by taking the value for the
gap estimated by other techniques and check if there is an agreement. We have 
fit the experimental data for $^{63}$Cu(2) (Fig. 3b) and $^{63}$Cu(1) (Fig.3a) 
in Sr$_{14}$Cu$_{24}$O$_{41}$
by taking $\Delta=450$ K and $\Delta=120$ K, respectively, as estimated
from susceptibility or NMR shift data \cite{Taki2,Car1}. In both cases we find a good agreement between
theory and experiment by taking $1\geq r\geq 0.5$ for the ladder site and $r\gg 1$ for the
chain site. If the data for $^{63}$Cu(2)
were fitted according to Eq. 2 one would
derive a value for the gap around $650$ K, a factor $1.5$ larger than the actual value 
(see Tab. 1).

 %Inserted by TeXtelmExtel

For $r= 1$ also a quantitative agreement with the experimental data for $^{63}$Cu(2)
is found. However,  this fact
seems to be at variance with the estimates by Johnston \cite{John} based on the analysis of
DC susceptibility data and with the recent findings by Imai et al. \cite{Imai} based on the
study of $^{17}$O NMR shift anisotropy, where  a value for $r\simeq 0.5$ was derived. 
If we take this value for
$r$ we find that the experimental data are a factor $\simeq 8$ larger than expected.
This disagreement could originate, at least partially, 
from having considered for the $q_x=2k_x$ processes the values for the 
$\vert<-k_x\vert S_{z}\vert k_x>\vert^2$ matrix elements estimated by Troyer for the case
$r=1$ \cite{Troyer}.
One has also to mention that the estimate of the hyperfine coupling constants could suffer from
some uncertainties, particularly the contribution from the 
transferred hyperfine interaction with the neighbouring Cu$^{2+}$ spins. 
This contribution should be particularly
relevant for the $^{63}$Cu(1) nuclei while it should be small for $^{63}$Cu(2). However, it
must be recalled that since  $1/T_1$ depends quadratically on the hyperfine coupling constant
even for $^{63}$Cu(2) sizeable corrections can be exepected. 
Finally it has to be observed that in these systems the low-frequency divergence of $1/T_1$
is cut because of the finite coupling among the ladders (or chains), introducing another
correction to the absolute value of $1/T_1$.

 %Inserted by TeXtelmExtel

The low-frequency divergence of $1/T_1$ was found to follow the logarithmic behavior reported
by Troyer et al. \cite{Troyer} (see also Eq. 2) and 
does not change upon varying the anisotropy factor
$r$, for  $r>0$. In fact, the form of this divergence
is related to the shape of the dispersion curve
close to $k_x=\pi/a$, where it is always correctly approximated by a quadratic form for $r>0$.

 %Inserted by TeXtelmExtel

 %Inserted by TeXtelmExtel

In conclusion we have presented a careful analysis of the problem of estimating the spin-gap
from nuclear spin-lattice relaxation measurements in 1DQHAF. It is found that in order to
estimate correctly the gap one should either perform the experiments at temperatures $T\lesssim
0.2\Delta$ where in many cases other contributions to the relaxation process emerge 
\cite{Azuma,Taki2,Car1}, or use
an appropriate expression for $1/T_1$  which takes into account the form of the
dispersion curve for the triplet excitations. 
Then a good agreement for the gap value estimated by means 
of $1/T_1$ and other techniques is found, allowing also to derive information on the anisotropy
of the superexchange constants.  
 
 %Inserted by TeXtelmExtel

 %Inserted by TeXtelmExtel
 
We would like to thank D. C. Johnston for useful discussions. 
The research was carried out with the financial 
support of INFM and of INFN. 

 %Inserted by TeXtelmExtel

 %Inserted by TeXtelmExtel

 %Inserted by TeXtelmExtel

 %Inserted by TeXtelmExtel

%************TABLE 1*****************
\begin{table}
%\vspace{6.5cm}
\caption{Values for the gap $\Delta$ between singlet and triplet excitations for different
1DQHAF, estimated from $1/T_1$ using Eq. 2 and from DC susceptibility (or NMR shift)
measurements. In the last column the ratio for the values of the gap estimated by the two
techniques is reported.}
\end{table}

 %Inserted by TeXtelmExtel

%************FIGURE 1*******************
\begin{figure}
%\vskip 6.5cm
\caption{The dispersion curves for the triplet excitations in a 2-leg-ladder are reported for 
different values of the superexchange anisotropy $r=J_{\perp}/J_{\parallel}$. 
The dotted line
shows the quadratic approximation of the dispersion curve for $r>>1$ (see Eq. 1).
}
\end{figure}

 %Inserted by TeXtelmExtel

 %Inserted by TeXtelmExtel

 %Inserted by TeXtelmExtel

 %Inserted by TeXtelmExtel

%************FIGURE 2*****************
\begin{figure}
%\vspace{6.5cm}
\caption{a) Nuclear spin-lattice relaxation rate as a function of $\Delta/T$ for three different
values of the ratio $r=J_{\perp}/J_{\parallel}$. The solid lines give the results obtained from Eqs.
3 and 4 after a numerical integration for $\omega_o/2\pi=15$ MHz, while the dotted lines show the corresponding
behavior by using the quadratic approximation (see Eq. 2). b) The same data as in a) are 
now reported as a function of $J_{\parallel}/T$. }
\end{figure}

 %Inserted by TeXtelmExtel

 %Inserted by TeXtelmExtel

 %Inserted by TeXtelmExtel

 %Inserted by TeXtelmExtel

 %Inserted by TeXtelmExtel

%************FIGURE 3******************
\begin{figure}
%\vskip 6.5cm
\caption{a)Temperature dependence of $^{63}$Cu(1) $1/T_1$  in a Sr$_{14}$Cu$_{24}$O$_{41}$
single crystal 
for $\vec H \parallel b$. The solid line shows the behavior expected from Eq. 3 by taking
a gap $\Delta= 120$ K. 
b)Temperature dependence of $^{63}$Cu(2) $1/T_1$ in Sr$_{14}$Cu$_{24}$O$_{41}$ for $\vec H\parallel
b$. The data were obtained either in oriented powders (circles) or single crystals (squares).
The solid line shows the behavior according to Eq. 3 for $r=0.5$ by using the same value 
for the gap derived
from $^{63}$Cu(2) NMR shift ($\Delta= 450$ K), 
while the dotted line gives the corresponding behavior obtained using the
quadratic approximation (Eq. 2) with $\Delta=650$ K.}
\end{figure}

 %Inserted by TeXtelmExtel

 %Inserted by TeXtelmExtel

 %Inserted by TeXtelmExtel

 %Inserted by TeXtelmExtel

 %Inserted by TeXtelmExtel

\newpage
.

 %Inserted by TeXtelmExtel

 %Inserted by TeXtelmExtel

 %Inserted by TeXtelmExtel

 %Inserted by TeXtelmExtel


\begin{references}

 %Inserted by TeXtelmExtel

\bibitem{Dagotto} E. Dagotto and T. M. Rice, Science 271, 619 (1995)

 %Inserted by TeXtelmExtel


 %Inserted by TeXtelmExtel

\bibitem{Uehara} M. Uehara et al., J. Phys. Soc. Jpn. 65, 2764 (1996)

 %Inserted by TeXtelmExtel

\bibitem{Tranquada} J. M. Tranquada et al., Nature 375, 561 (1995) 

 %Inserted by TeXtelmExtel

\bibitem{Car0} P. Carretta, Physica C 292, 286 (1997)

 %Inserted by TeXtelmExtel

\bibitem{Azuma} M. Azuma et al., Phys. Rev. Lett. 73, 3463 (1994)

 %Inserted by TeXtelmExtel

\bibitem{Taki1} M. Takigawa et al., Phys. Rev. Lett. 76, 2173 (1996)

 %Inserted by TeXtelmExtel

\bibitem{Taki2} M. Takigawa et al., Phys. Rev. B 57, 1124 (1998)

 %Inserted by TeXtelmExtel

\bibitem{Kumagai} K. Kumagai et al., Phys. Rev. Lett. 78, 1992 (1997)

 %Inserted by TeXtelmExtel

\bibitem{Magishi} K. Magishi et al., Phys. Rev. B 57, 11533 (1998)

 %Inserted by TeXtelmExtel

\bibitem{Car1} P. Carretta et al., Phys. Rev. B 56, 14587 (1997)

 %Inserted by TeXtelmExtel

\bibitem{Car2} P. Carretta, A. Vietkin and A. Revcolevschi, Phys. Rev. B 57, R5606 (1998)

 %Inserted by TeXtelmExtel

\bibitem{Mayaf} H. Mayaffre et al., Science 279, 345 (1998)

 %Inserted by TeXtelmExtel

\bibitem{Tedo} F. Tedoldi et al., J. App. Phys. 83, 6605 (1998)

 %Inserted by TeXtelmExtel

\bibitem{Imai} T. Imai et al., Phys. Rev. Lett. 80, (1998)

 %Inserted by TeXtelmExtel

\bibitem{Greg} G. Chaboussant et al., Phys. Rev. Lett. 79, 925 (1997)

 %Inserted by TeXtelmExtel

\bibitem{Furuk} Y. Furukawa et al., J. Phys. Soc. Jpn. 65, 2393 (1996)

 %Inserted by TeXtelmExtel

\bibitem{Troyer} M. Troyer, H. Tsunetsugu and D. W\"urtz, Phys. Rev. B 50, 13515 (1994)

 %Inserted by TeXtelmExtel

\bibitem{Damle} S. Sachdev and K. Damle, Phys. Rev. Lett. 78, 943 (1997)

 %Inserted by TeXtelmExtel

\bibitem{Strut} E. M. McCarron et al., Mater. Res. Bull. 23, 1355 (1988)

 %Inserted by TeXtelmExtel

\bibitem{Oitmaa} J. Oitmaa, R. R. P. Singh and Z. Weihong, Phys. Rev. B 54, 1009 (1996)

 %Inserted by TeXtelmExtel

\bibitem{John} D. C. Johnston, Phys. Rev. B 54, 13009 (1996)

 %Inserted by TeXtelmExtel


 %Inserted by TeXtelmExtel


 %Inserted by TeXtelmExtel


 %Inserted by TeXtelmExtel


 %Inserted by TeXtelmExtel


 %Inserted by TeXtelmExtel


 %Inserted by TeXtelmExtel

\end{references}
\end{document}